\documentclass[preprint]{elsarticle}
\makeatletter
\def\ps@pprintTitle{%
	\let\@oddhead\@empty
	\let\@evenhead\@empty
	\def\@oddfoot{}%
	\let\@evenfoot\@oddfoot}
\makeatother
\usepackage[colorlinks = true,
linkcolor = blue,
urlcolor  = blue,
citecolor = blue,
anchorcolor = blue]{hyperref}
\usepackage{balance}
\usepackage{amsmath}
\usepackage{amssymb}
\usepackage{mathrsfs}
\usepackage{graphicx}
\usepackage{epstopdf}
\usepackage{bm}
\usepackage{color}
\usepackage{subfig}

\begin{document}

\begin{frontmatter}

\title{How many can you infect? \\Simple (and naive) methods of estimating the reproduction number}
\tnotetext[mytitlenote]{Corresponding author. Email Address: hadi.susanto@ku.ac.ae (H.\ Susanto).}

\author[a,b]{H.\ Susanto *}
\author[c]{V.R.\ Tjahjono}
\author[d]{A.\ Hasan}
\author[e]{M.F.\ Kasim}
\author[f]{N.\ Nuraini}
\author[c]{E.R.M.\ Putri}
\author[f]{R.\ Kusdiantara}
\author[c]{H.\ Kurniawan}

\address[a]{Department of Mathematics, College of Arts and Sciences, Khalifa University,\protect\\PO Box 127788, Abu Dhabi, United Arab Emirates}
\address[b]{Department of Mathematical Sciences, University of Essex, \protect\\Wivenhoe Park, Colchester, CO4 3SQ, United Kingdom}
\address[c]{Department of Mathematics, Faculty of Mathematics and Natural Sciences, Institut Teknologi Sepuluh Nopember, Sukolilo, Surabaya 60111, Indonesia}
\address[d]{Center for Unmanned Aircraft Systems M\ae rsk McKinney M\o ller Institute, \protect\\University of Southern Denmark, 5230 Odense, Denmark}
\address[e]{Clarendon Laboratory, Department of Physics, University of Oxford,\protect\\ Parks Road, Oxford, United Kingdom}
\address[f]{Industrial and Financial Mathematics Research Group, Department of Mathematics, \protect\\Institut Teknologi Bandung, Ganesha 10, Bandung, 40132, Indonesia}

\begin{abstract}
This is a pedagogical paper on estimating the number of people that can be infected by one infectious person during an epidemic outbreak, known as the reproduction number. Knowing the number is crucial for developing policy responses. There are generally two types of such a number, i.e., basic and effective (or instantaneous). While basic reproduction number is the average expected number of cases directly generated by one case in a population where all individuals are susceptible, 
effective reproduction number is the number of cases generated in the current state of a population. In this paper, we exploit the deterministic susceptible-infected-removed (SIR) model to estimate them through three different numerical approximations. We apply the methods to the pandemic COVID-19 in Italy to provide insights into the spread of the disease in the country. 
We see that the effect of the national lockdown in slowing down the disease exponential growth appeared about two weeks after the implementation date. We also discuss available improvements to the simple (and naive) methods that have been made by researchers in the field. 

Authors of this paper are members of the SimcovID (Simulasi dan Pemodelan COVID-19 Indonesia) collaboration. 
\end{abstract}

\begin{keyword}
Reproduction number, infectious disease, compartment model, and COVID-19
\end{keyword}

\end{frontmatter}


\section{Introduction}
\label{sec:intro}


\emph{"When will the peak of the pandemic hit? When will it be over?"}

Those are unarguably among the most asked questions during the ongoing coronavirus disease 2019 (COVID-19) crisis, i.e., a disease outbreak of atypical pneumonia that originated from Wuhan, China \cite{ncov20}. The disease spread to over 100 countries in a matter of weeks \cite{call20}, with most internationally imported cases reported to date having history of travel to Wuhan \cite{wu20,kuch20}. The pandemic has made governments all over the world take serious responses \cite{wiki20}. The governmental measures result in a significant disruption in the lives of their people that raised such questions above.

Diseases grow rather exponentially at the initial transmissibility of outbreak \cite{ma20,chow16}. When there is no intervention and the proportion of infections starts to become comparable to the entire population, the growth will slow down as susceptible is fuel to diseases. This type of logistic growth will yield the peak of a pandemic that everybody is interested in and its arrival can be forecast using, e.g., the susceptible-infected-removed (SIR) compartment model \cite{ali14}. 

However, in the presence of pandemic, human beings adapt. Governments intervene. As such, using the SIR model to predict the peak, while new cases are mainly outcomes of national policies and/or community behaviour, would be similar to forecasting what policymakers would do or the effectiveness of their response \cite{gume04,yang20,fang20}, which is dynamic and can be unprecedented. On top of that, there is a lack of knowledge of epidemiology characteristics and 
a high rate of undocumented cases \cite{li20}. A brute force analysis by fitting reported data to the SIR model is therefore prone to a false prediction if not done carefully (see, e.g., Fig.\ 2 of \cite{fane20} that incorrectly predicted the peak time as well as the total infection of COVID-19 in Italy when compared to the latest data). We will show below how data-driven forecasts are sensitive to the time-series information that we input in the model. 

Considering the limitations and obstacles, it is therefore important to determine instead the so-called disease reproduction number or reproductive factor \cite{diet93}, which is the number of people that are infected by one infectious person during an epidemic outbreak \cite{chow09,nish09,drie17,ride18,dela19}. It depends on the duration of the infectious period, the probability of infecting a susceptible individual during one contact, and the number of susceptible people contacted per unit of time. 

There are generally two types of such a number, i.e., basic \cite{chow09} and effective (or instantaneous) \cite{nish09}. While basic reproduction number is the average expected number of cases directly generated by one case in a population where all individuals are susceptible, effective reproduction number is the number of cases generated in the current state of a population. This paper is intended to give a brief review of these numbers to undergraduate students 
and a broad science-educated audience in general. We also hope that the paper can be an expository article of epidemiology to policyholders in making public health measures. 

To quantify directly the actual reproduction number is difficult, if not impossible, and as such, we can only estimate it indirectly. One common approach is to fit a model to epidemiological data that will provide values of some parameters \cite{ma20}. Here, we use the SIR compartment model as our model reference, where the reproduction number is associated to the threshold point for stability of the disease free equilibrium. 

There are three estimation methods that we will discuss. As a case study, we apply the methods to discuss and forecast viral transmission of COVID-19 in Italy. The first one is by parameter fit to the SIR model \cite{cint09}, which is probably the most popular analysis to the study of COVID-19 \cite{fane20}. The computed parameters will then be used to obtain the reproduction number. 
The second method is to use the reported data of infected and removed people \cite{chen20}. Comparing the number with that obtained using the parameter fit shows a similar trend in the decrease of the infection rate in Italy. The third method is using the ratio of increment of infections from two subsequent days \cite{bett08,chow07}. However, such a quantity is usually highly fluctuating as we demonstrate for the case of Italy. The trend is obtained using, e.g., a parameter fit of the Richards curve \cite{ma20,nura20} to the cumulative cases. 

As the methods presented here are all based on the SIR model, they are limited by assumptions commonly made within the SIR model. An important assumption is that the presented data are an accurate representation of what happens in the population, although this can be relaxed for some methods in this paper. Another assumption or limitation is that it does not include people that are infected but not infectious, which can be overcome by incorporating another compartment, such as the commonly used Exposed group.

We conclude the paper with a brief review of improvements to the methods by including randomness (stochastic processes).



\section{SIR model and the reproduction number}
\label{sec2}

The SIR model equations are given by 
\begin{align}
&\frac{dS}{dt}=-\beta\frac{SI}{N},\label{sir1}\\
&\frac{dI}{dt}=\beta\frac{SI}{N}-\gamma I,\label{sir2}\\
&\frac{dR}{dt}=\gamma I.\label{sir3}
\end{align}
Here, $S$ and $I$ denote the total number of susceptible and infected individuals. Variable $R$ represents the removed compartment that can consist of recovered (and become-resistant) and deceased individuals. The total population is $N=S+I+R$. Note that $dN/dt=d(S+I+R)/dt=0$, which implies that $N$ is constant. The parameters $\beta$ and $\gamma$ are the transmission and removal rate constants, respectively. The average length of time an infected individual remains infective, i.e., the infectious time, is $1/\gamma$. Note that this still applies even when the parameters $\beta$ and $\gamma$ are functions of time. Additionally, we denote the cumulative (total) case as $T=I+R$, which satisfies the equation 
\begin{equation}
\frac{dT}{dt}=\beta \frac{SI}{N}.
\label{T}
\end{equation}

Equation \eqref{sir2} can also be written as 
\begin{equation}
\frac{dI}{dt}=\gamma\left(\mathcal{R}_t-1\right)I,
\label{t1}
\end{equation}
where 
\begin{equation}
\mathcal{R}_t=\frac{S}{N}\mathcal{R}_0,\quad \mathcal{R}_0=\beta/\gamma.
\label{R}
\end{equation}
$\mathcal{R}_t$ is the effective reproduction number and $\mathcal{R}_0$ is the basic one. Note from \eqref{t1} that depending on the value of $\mathcal{R}_t$, i.e., whether $\mathcal{R}_t>1$ or $\mathcal{R}_t<1$, the infections $I$ will increase or decrease in time, respectively. It is therefore important to track this number to forecast the spread of an infection in an area.

As data are collected and reported regularly in a certain time interval, it is instructive to consider instead the discrete model
\begin{align}
&\Delta S_{n}=-\tau\beta\frac{S_nI_n}{N},\label{dsir1}\\
&\Delta I_{n}=\tau\beta\frac{S_nI_n}{N}-\tau\gamma I_n,\label{dsir2}\\
&\Delta R_{n}=\tau\gamma I_n,\label{dsir3}
\end{align}
where $\Delta K_n=K_{n+1}-K_{n}$, $K=S,\,I,\,R$. $\tau$ is the time interval, which in the limit $\tau\to0$, make the model \eqref{dsir1}-\eqref{dsir3} become \eqref{sir1}-\eqref{sir3}. We take $\tau=1$ day, which is the standard time interval to report updates on cases for COVID-19. The reproduction numbers \eqref{R} can be checked to be still applicable here. In the following, we will mainly use the discrete SIR model \eqref{dsir1}-\eqref{dsir3}.

As the main part of this report, we will consider three different methods to approximate the effective reproduction number $\mathcal{R}_t$. 

\subsection{Method 1: Parameter fit}

\begin{figure}[tbhp]
	\centering
\subfloat[]{\includegraphics[width=\linewidth]{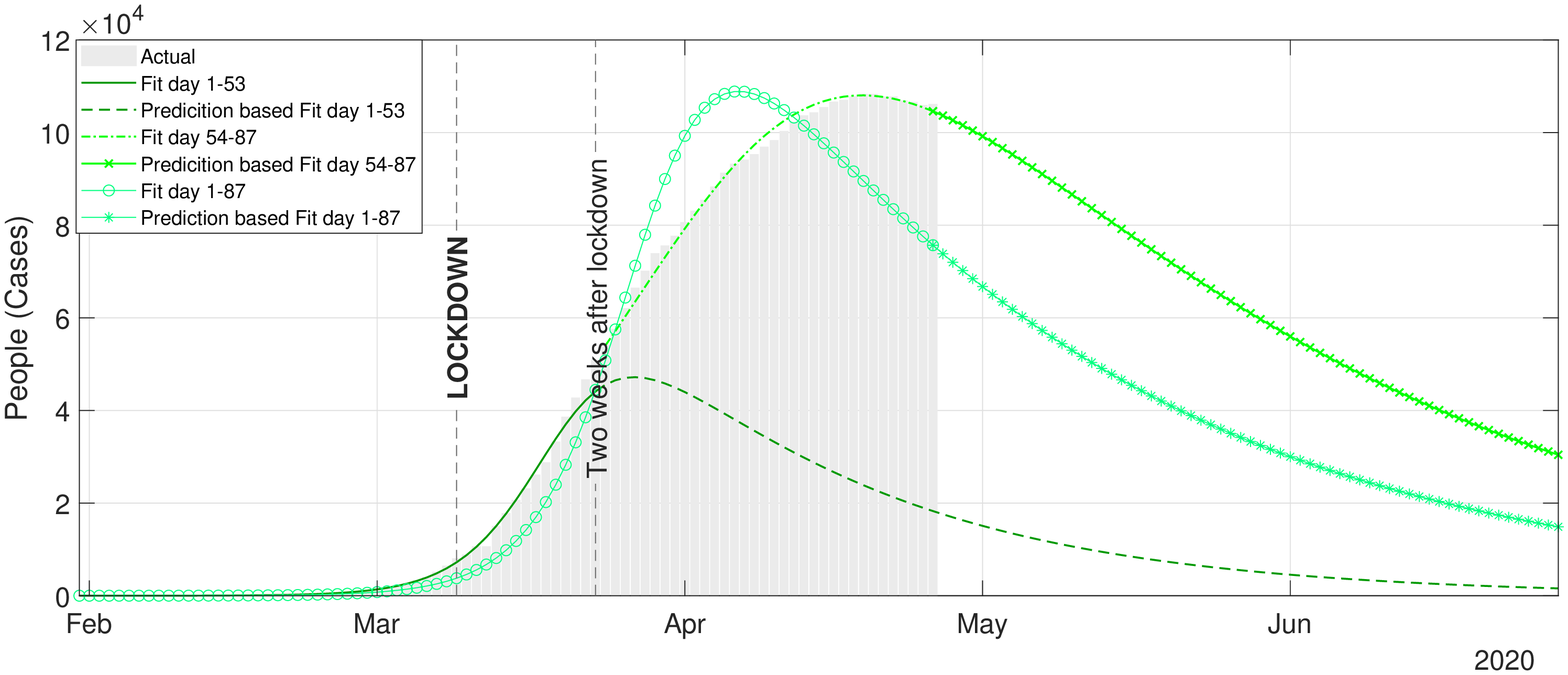}}\\
\subfloat[]{\includegraphics[width=\linewidth]{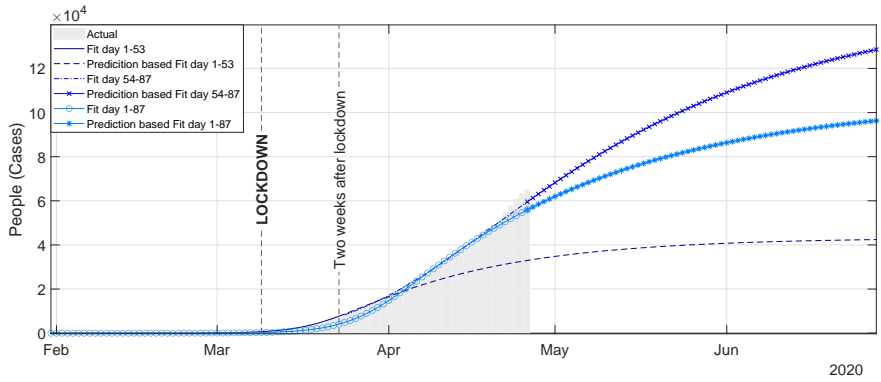}}\\
\subfloat[]{\includegraphics[width=\linewidth]{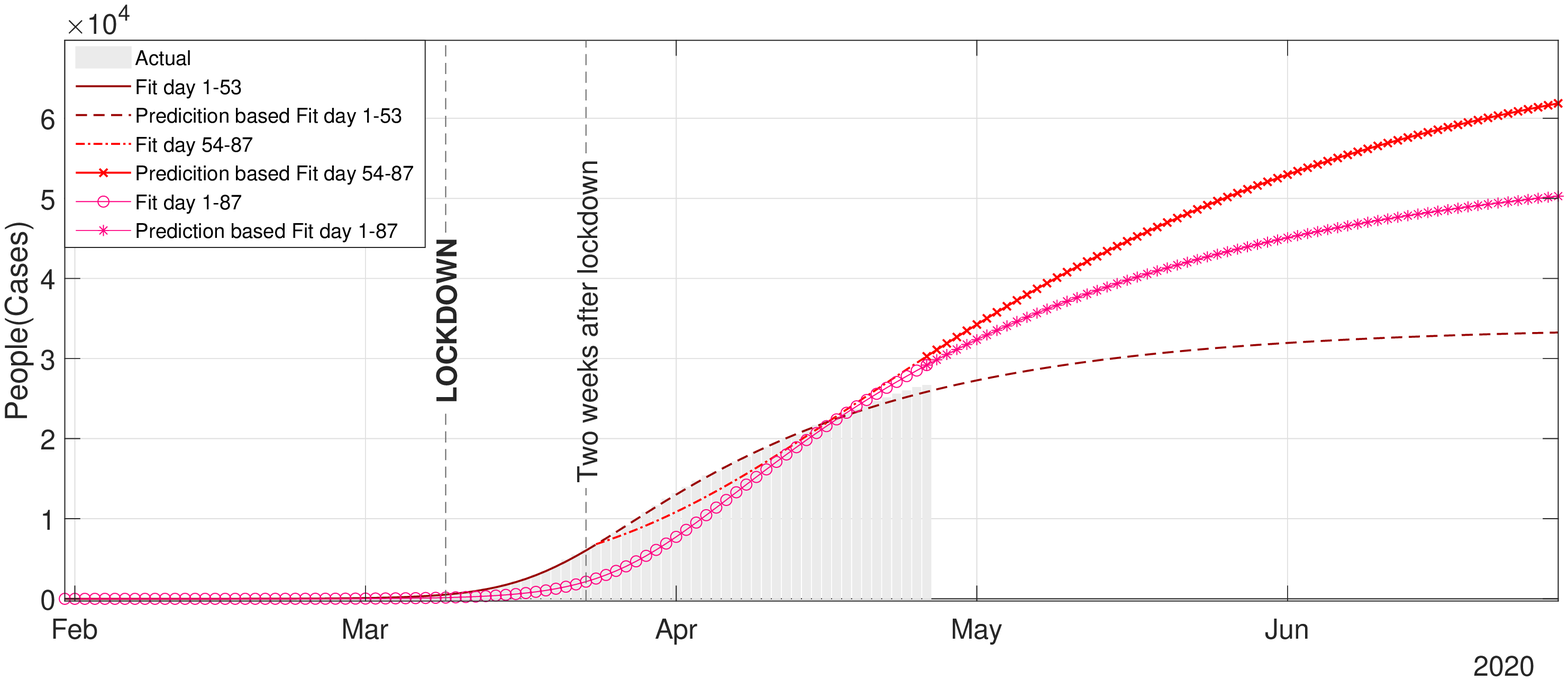}}
	\caption{Predicted evolution of the COVID-19 outbreak in Italy based on the fits of the discrete SIR model \eqref{dsir1}-\eqref{dsir3} for (a) active cases; (b) recovered; (c) deceased. Shaded regions represent the official data retrieved from the JHU CSSE repository \cite{data}. There are three different predicted trends, based on the length of the fitted data, see the legends. Day 1 = 31 Jan 2020, Day 39 (national lockdown) = 9 Mar 2020, Day 87 = 26 Apr 2020.}
	\label{fig1}
\end{figure}

To calculate $\mathcal{R}_t$ \eqref{R}, one needs to determine first the parameters $\beta$ and $\gamma$, as well as the number of susceptible $S_n$ and hence the population size $N$. Because infection data are given in terms of the number of infected and removed (i.e., recovered or deceased) people, we can find the parameter set for which the model has the best agreement with the data. In that case, we fit the deterministic epidemiological model \eqref{dsir1}-\eqref{dsir3} by employing a generalized least squares scheme, i.e., we search for the minimum of an unconstrained problem specified by
\begin{equation}
\displaystyle\min_{\{S_1,\beta,\gamma\}}\sum_n\left(I_n-Idata_n\right)^2+\left(R_n-Rdata_n\right)^2,
\end{equation}
where $Idata_n$ and $Rdata_n$ are reported infected and removed cases at day $n$. Here we only limit ourselves to minimization using three parameters ($S_1$, $\beta$, and $\gamma$) only. Note that $N$ is implicitly part of the estimated parameters because $N=S_1+I_1+R_1$. We take $I_1=Idata_1$ and $R_1=Rdata_1$ at the initial step. 

The search is done using \textbf{fminsearch} function of \textsc{Matlab} that implements the Simplex search method. To illustrate our computation, we consider COVID-19 cases in Italy, which was one of the world's worst-hit countries. Data were retrieved from \cite{data} on 26 April 2020, which in the analysis will be denoted as Day 87 (i.e., Day 1 is 31 Jan 2020). We present in Fig.\ \ref{fig1} the reported data and the fitting SIR dynamics. We slightly modify the model by splitting the removed compartment ($R$) into recovered and deceased ones with rates $\gamma^{(1)}$ and $\gamma^{(2)}$ respectively and as such, $\gamma=\gamma^{(1)}+\gamma^{(2)}$.

\begin{table}[t]
	\begin{center}
		\begin{tabular}{c|c|c|c} 
			& Fitted Day 1-53 & Fitted Day 53-87 & Fitted Day 1-87\\
			\hline
			$S_1$ & 77301.031 & 154875.683 & 161479.047\\
			$\beta$ & 0.282 & 0.119 & 0.246\\
			$\gamma^{(1)}$ & 0.021 & 0.0169  & 0.017\\
			$\gamma^{(2)}$ & 0.017& 0.008 & 0.009\\
		\end{tabular}
	\end{center}
	\caption{Parameters obtained from the minimization procedure. }
	\label{tab:table1}
\end{table}

From our computations, using all the available data, i.e., Day 1-87, to be fitted into the SIR equations yields rather bad agreement. It turned out to be necessary to split the data into two periods, separated around the national lockdown that was implemented on 9 March 2020 (Day 39). To be precise, it is the threshold date of the lockdown effect that manifested two weeks later, i.e., from Day 53. 

The split is to incorporate the intervention and behavioural change of the population in the model that requires the parameter values to vary over time. Note that the SIR model assumes the parameter values to be constant during the fitted period, which is not necessarily correct. Assuming constant value of those parameters implicitly assumes that the decline in active cases is because herd immunity (i.e., substantial decline in susceptible population) has been formed, which has not been detected anywhere, even at places with high death counts. 
Splitting the graph and fitting the parameters separately are therefore to solve the assumption violation, where an extra care must be taken in the procedure.

Using the splitting, we obtain good agreement as can be seen in Fig.\ \ref{fig1}. It is important to note that using data from Day 1-53, we obtained a predicted peak at the end of March 2020, which clearly is not correct, i.e., parameter fits depend sensitively on the fitted data. This explains the incorrect prediction of \cite{fane20}.

In Table \ref{tab:table1}, we list the fitting parameters. Using the values, we plot in Fig.\ \ref{fig2} the resulting estimated effective reproduction number $\mathcal{R}_t$ \eqref{R}. It is clear that the national lockdown effectively decreased the number. The curve crosses the axis $\mathcal{R}_t=1$ at the peak on 19 Apr 2020. 

\subsection{Method 2: Using infected and recovered data}

\begin{figure}[tb!]
	\centering
	\includegraphics[width=\linewidth]{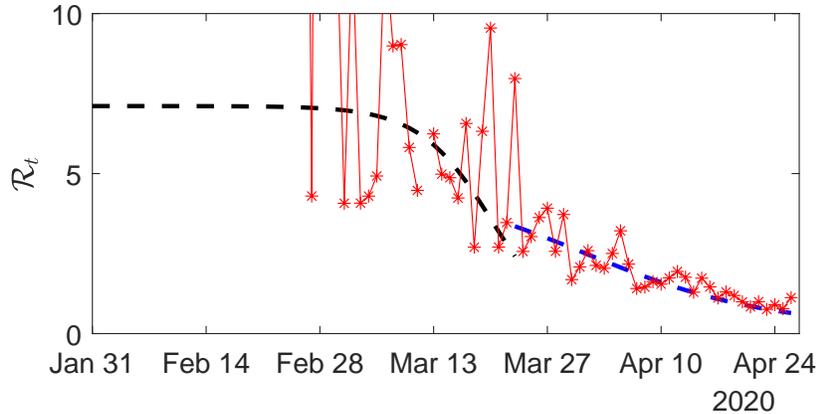}
	\caption{Estimated effective reproduction number using Method 1 (dashed lines) and Method 2 (stars). The lines are necessarily split into two parts following the national lockdown, see the text.}
	\label{fig2}
\end{figure}

In the second method, instead of getting the parameters $\beta$ and $\gamma$ from fitting, we will derive them from the governing equations \eqref{dsir2}-\eqref{dsir3} directly. Writing $\beta\equiv\beta_n$ and $\gamma\equiv\gamma_n$ on the right hand side of the equations, it is straightforward to obtain
\begin{equation}
\beta_n=\frac{\Delta(I_n+R_n)}{\tau S_nI_n}N,\quad \gamma_n=\frac{\Delta R_n}{\tau I_n}.
\label{bg}
\end{equation}

From the definition \eqref{R}, we have $\mathcal{R}_0=\beta_n/\gamma_n$ and as such, 
\begin{equation}
\mathcal{R}_t=\frac{S_n}N\mathcal{R}_0=1+\frac{\Delta I_n}{\Delta R_n}.
\end{equation}
We therefore obtain that the effective reproduction number is related to the ratio between the change of the infected and the removed compartments. Because $\Delta R_n>0$, then $\mathcal{R}_t<1$ if and only if $\Delta I_n<0$.

We show in Fig.\ \ref{fig2} the estimated reproduction number using the second method, depicted in stars. Because it uses the increase in infected and removed counts which tend to be highly fluctuating, the resulting curve is also wavering. This could be simply solved by smoothing the data using a moving average filter. 
Nevertheless, in our case here, we still can observe that it is following the same trend as that obtained from Method 1, i.e., the dashed curve. 


\subsection{Method 3: Using new cases}

\begin{figure}[tbhp!]
	\centering
	\subfloat[]{\includegraphics[width=\linewidth]{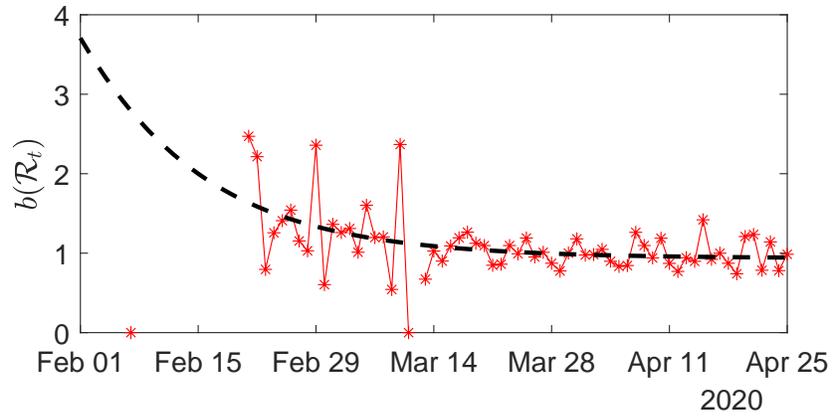}}\\	\subfloat[]{\includegraphics[width=\linewidth]{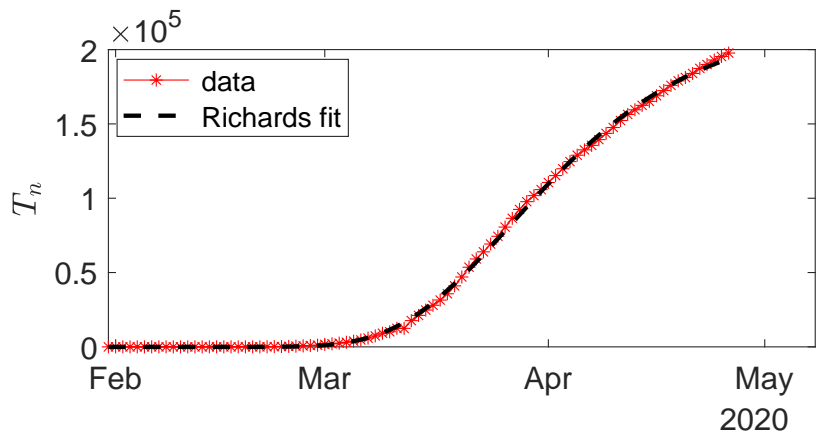}}
	\caption{(a) Plot of $b(\mathcal{R}_t)$ in time from the reported data (stars) and the approximation obtained using Richards' curve (dashed line). (b) Cumulative cases from data (stars) and Richards' approximation (dashed line).}
	\label{fig3}
\end{figure}

The third method is to exploit the daily reported new cases, which in terms of the SIR model will be given by the daily difference of the cumulative cases $\Delta T_n$. Integrating \eqref{sir2} in time between $t$ and $(t+\tau)$ gives us
\begin{equation}
\displaystyle I_{n+1}=I_ne^{\gamma\int_t^{t+\tau}(\mathcal{R}_t-1)dt}\simeq I_nb(\mathcal{R}_t),
\end{equation}
where
\begin{equation}
b(\mathcal{R}_t)=e^{\gamma\tau(\mathcal{R}_t-1)}.
\label{brt}
\end{equation} 
Here, we denote $I(t+\tau)=I_{n+1}$ and $I(t)=I_n$. In the last equation, we have assumed that $\mathcal{R}_t$ is constant within the time interval. 

On the other hand, we have from \eqref{T} a discrete approximation
\begin{equation}
\Delta T_n=\tau\beta{S_{n+1}I_{n+1}}/{N}=\tau\beta{S_{n+1}I_nb(\mathcal{R}_t)}/{N}
\simeq\tau\beta{S_{n}I_nb(\mathcal{R}_t)}/{N}.
\label{temp}
\end{equation}
The last step is expected to be valid for emerging diseases, i.e., $I$ varies slowly.

At the same time, we also have from \eqref{T}
\begin{equation}
\Delta T_{n-1}=\tau\beta{S_{n}I_{n}}/{N}.
\label{temp1}
\end{equation}

Combining \eqref{temp} and \eqref{temp1} gives us the effective reproduction number
\begin{equation}
b(\mathcal{R}_t)=\Delta T_{n}/\Delta T_{n-1},\quad \mathcal{R}_t=1+\frac{1}{\tau\gamma}\ln\left(b(\mathcal{R}_t)\right).
\end{equation}
Because $b(\mathcal{R}_t)$ is a monotonically increasing function in $\mathcal{R}_t$, it can be enough to plot $b$ itself to determine whether the disease decreases or not. 

In Fig.\ \ref{fig3}(a) we plot $b(\mathcal{R}_t)$ from the data of COVID-19 cumulative cases in Italy shown in stars. However, the curve is highly fluctuating that may hide the trend. As mentioned in Method 2 above, this could also be solved by smoothing the data using a moving average filter. 
Another possible way 
is by approximating the reported cumulative cases with a continuous function. A natural candidate is certainly the generalised logistic function, also known as Richards' curve,
\begin{equation}
T_n=\frac{A}{\left(B+e^{-C(n-n_0)}\right)^{1/\nu}}.
\label{Tn}
\end{equation}
Again using a least square method to fit the reported cumulative cases to the function, we obtain that the best parameter values are 
$A=119520$, $B=0.9858$, $C=0.0665$, $n_0=0.0086$, $\nu=0.0232$. We plot the fitted data and the approximation in Fig.\ \ref{fig3}(b) in stars and dashed line, respectively. 

Using the approximation \eqref{Tn}, we are now able to plot a smooth curve (shown in dashed in Fig.\ \ref{fig3}(a)) that is expected to indicate the trend of $b(\mathcal{R}_t)$ calculated from the reported data.


\section{Conclusion}

We have presented three simple (or actually simplistic) methods to estimate the reproduction number of the COVID-19 pandemic based on the SIR equations as the underlying model. We applied the methods to the data of COVID-19 cases in Italy, where we saw that the implemented national lockdown had positive impacts that appeared about two weeks later.  

To extend the deterministic methods reviewed herein, one may consider complex models that include more compartments \cite{ma20,li11}. However, to be more realistic, one should include statistical randomness and probability in the calculations. 

In the spirit of Method 1, Cintr\'on-Arias et al.\ \cite{cint09} combined parameter fits with statistical asymptotic theory and sensitivity analysis to give approximate sampling distributions for the estimated parameters. Method 3 has been improved in \cite{bett08,chow07} to include a probabilistic description such that 
the probabilistic formulation for future cases is equivalent, via Bayes’ theorem, to the estimation of the probability distribution for the reproduction number. 

In addition to estimating the reproduction number based on a model, it is also possible to approximate the reproduction number from the serial interval (the time between the onset of symptoms in a primary case and the onset of those in secondary cases) without assuming a model 
\cite{cori13,wall04,obad12}.

\section*{Acknowledgement}

H.S.\ is extremely grateful to his wife, dr.\ Nurismawati Machfira, who has happily taken a new additional job as 'head teacher' of their children at home during school closure, while maintaining her job as their primary carer, so that he could still \#workfromhome and wrote this paper. Part of this research is funded by Program Pengabdian Masyarakat ITB 2020.


\end{document}